\documentclass[9pt]{arxiv}

\templatetype{pnasmathematics}

\usepackage{epsfig,latexsym,cancel,amssymb,amsmath,bm}
\usepackage{graphicx}
\usepackage{feynmp}
\usepackage{subfigure}
\usepackage{epstopdf}
\usepackage{color}
\usepackage{mathtools}
\usepackage{bigints}
\usepackage[utf8]{inputenc}
\usepackage[english]{babel}

\usepackage[font=small,labelfont=bf]{caption}

\setboolean{displaywatermark}{false}

\graphicspath{{figs/}}

\title{Controlling moving interfaces in solid state batteries} 

\author[a, b]{Salem Mosleh}
\author[c]{Emil Annevelink} 
\author[c, d, e]{Venkatasubramanian Viswanathan}
\author[a, f, g, 1]{L. Mahadevan}

\affil[a]{John A. Paulson School of Engineering and Applied Sciences, Harvard University, Cambridge, MA 02138, USA}
\affil[b]{Department of Natural Sciences, University of Maryland Eastern Shore, Princess Anne, MD, 21853, USA}
\affil[c]{Department of Mechanical Engineering, Carnegie Mellon University, Pittsburgh, PA, 15213, USA}
\affil[d]{Wilton E. Scott Institute for Energy Innovation, Carnegie Mellon University, Pittsburgh, PA, 15213, USA}
\affil[e]{Department of Aerospace Engineering, University of Michigan, Ann Arbor, MI, 48109, USA}
\affil[f]{Department of Organismic and Evolutionary Biology, Harvard University, Cambridge, MA 02138, USA}
\affil[g]{Department of Physics, Harvard University, Cambridge, MA 02138, USA}

\leadauthor{Mosleh}

\significancestatement{Deep decarbonization of transportation requires safe batteries with   increased energy and power densities. All solid-state lithium metal batteries are a key technology promising all three of these. However, all solid-state lithium metal batteries can rarely cycle more than 100 times before the solid-solid interfaces significantly degrade. Here we propose a new morphogenesis framework to unite previous research in solid-state battery degradation and introduce a phenomenological model to accelerate design of persistent solid-solid interfaces.}

\authorcontributions{V.V. and L.M. conceived the research, all authors designed the research. S. M. and L.M. created the theoretical model, S.M. carried out the calculations, S. M. and E. A. wrote the first draft, all authors edited the paper.}
\authordeclaration{V.V. is a technical consultant at QuantumScape Corporation.}
\correspondingauthor{\textsuperscript{1}To whom correspondence should be addressed. E-mail: lmahadev@g.harvard.edu}


\begin{abstract}
Safe, all-solid-state lithium metal batteries enable high energy density applications, but suffer from instabilities during operation that lead to rough interfaces between the metal and electrolyte and subsequently cause void formation and dendrite growth that degrades performance and safety. Inspired by the morphogenetic control of thin lamina such as tree leaves that robustly grow into flat shapes --- we propose a range of approaches to control lithium metal stripping and plating. To guide discovery of materials that will implement these feedback mechanisms, we develop a reduced order model that captures couplings between mechanics, interface growth, temperature, and electrochemical variables. 
We find that long-range feedback is required to achieve true interface stability, while approaches based on local feedback always eventually grow into rough interfaces. All together, our study provides the beginning of a practical framework for analyzing and designing stable electrochemical interfaces in terms of the mechanical properties and the physical chemistry that underlie their dynamics. 
\end{abstract}

\dates{This manuscript was compiled on \today}

\begin{document}

\maketitle



\dropcap{L}ithium metal batteries promise higher energy density, safety, and charging rates, potentially enabling the widespread adoption of electric vehicles, trucks, and aircraft that rely on these for power \cite{albertus2021challenges, pasta20202020}. A primary obstacle to robust functioning of these batteries is associated with instabilities at the interface between lithium and the solid electrolyte during charging/discharging cycles leading to uneven lithium stripping and deposition \cite{lu2022void, cao2020lithium}, which leads to the formation of interfacial voids and dendrites. These defects  can short-circuit the battery and cause failure \cite{liu2022unlocking}. 

Quenching the instability at the solid-solid interface has been the major focus of solid-state battery research for some time \cite{boarettoLithiumSolidstateBatteries2021}. When this fails,
voids form due to non-conformal contact between the solid-solid interface and grow from manufacturing inhomogeneities \cite{ishikawaCrystalOrientationDependence2017, shishvan2021initiation} when lithium is stripped from the anode and are exacerbated by local current concentration at void edges \cite{eckhardt3DImpedanceModeling2022}.
Throughout cycling, current density enhancement creates the conditions for filament formation, where narrow lithium deposits grow from the anode to the cathode creating internal shorts and ending the usable life of a battery \cite{bucci2017modeling}. On the other hand, maintaining a flat interface between the lithium electrode and the solid electrolyte can go a long way in reducing the propensity for battery failure.


Elucidating the role of electrochemical stability on the cycle life of all solid state batteries made it possible to create a design space of solid electrolytes and additives \cite{xuInterfacialChallengesProgress2018}, enabling improvements in processing such as ball milling to increase initial performance. However, enhancement in cycle life \cite{shinInterfacialArchitectureExtra2014} did not automatically follow. 
Over the last decade, two approaches have been proposed to enhance battery stability and robustness, increasing stack pressure and adding artificial interfaces. On its own, stack pressure cannot stabilize lithium deposition \cite{tuElectrodepositionMechanicalStability2020} and in some cases, might make the instability more prone to dendrite formation \cite{fincherControllingDendritePropagation2022}. However, 
due to the compliance of lithium metal, external pressure increases contact and simultaneously alters the interfacial kinetics to modulate and sometimes enhance interface stability \cite{ahmadStabilityElectrodepositionSolidSolid2017,barai2017lithium}.  In contrast,  creating artificial interfaces can either increase or decrease stability; using sputtered or vapor deposited layers of Aluminum or Germanium alloys increases wettability and decreases interfacial resistance \cite{fuGarnetElectrolyteBased2017,luoReducingInterfacialResistance2017}, while Silver-Carbon and Tungsten interfacial materials showed promising cyclability of solid-state batteries~~\cite{leeHighenergyLongcyclingAllsolidstate2020,raj2022direct}.
All together, it is clear that we need to consider a combination of global regulation using external stack pressure and local chemical modifications to quench interface instability by changing the chemical response to alter the stable operating regime of solid-state batteries.

Here, we suggest a framework to achieve stable interface motion inspired by the observations  of tightly regulated growth of living systems, morphogenesis, such as leaves that can maintain their flat shapes using various feedback mechanisms \cite{robustness-plants,plant-read-shape, 2021roadmap,flucts-pcepshn-sci}. The reproducible patterns seen in morphogenesis are associated with the coupling of mesoscopic biophysical processes to molecular genetics that together control the transport, deformation, and flows underlying growing structures \cite{thompson-growth&form, turingChemicalBasisMorphogenesis1952, mahadevan2020}. Just as bio-inspired closed-loop feedback mechanisms can stabilize form, we propose that stabilizing the interface between electrodes and electrolytes can prevent the formation of void and dendrites. Indeed, conceptualizing both electrochemical interface and biological growths as examples of morphogenesis allows insights to flow between the two domains and provides a unique vantage point from which to think about regulating electrochemical interfaces --- since their biological counter parts have evolved over millions of years to regulate shape with functional correlates that are subject to natural selection. 

\begin{figure*}[t]
    \centering
\includegraphics[width=\textwidth]{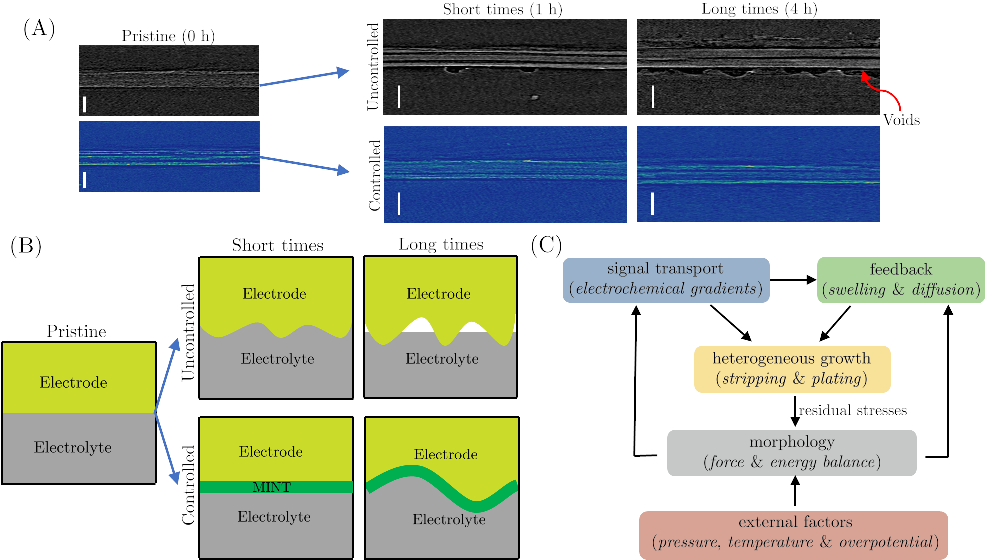}
\caption{\textbf{Preventing void formation with morphogenic interfaces}. (A) Cross-sectional analysis of synchrotron hard X-ray tomograms, visualizing both the advancing and receding electrode–electrolyte interfaces in Li–Li cells configured with standard separator (top) or a composite separator (bottom), after the cells had been polarized for different durations at 1 mA cm$^{-2}$. Scale bars, 50 $\mu$m. From ref. \cite{fu2020universal} Figure (B) Shows how the flat (pristine) interface between an electrode and electrolyte evolves under uncontrolled (normal) and controlled deposition, which involves the addition of a morphogenic interface (MINT) that responds to deviations from the flat state and changes how Lithium is deposited accordingly. Due to feedback the amplitude of interface undulations is smaller and smoother (longer wavelength) and avoids loss of contact. (C) A schematic illustrating morphogenesis (shape generation and regulation) in biological and electrochemical systems. Heterogeneous growth, in the form of stripping and plating in batteries, leads to mechanical, chemical, and thermal stresses that function as signals that further modify deposition. Adding the MINT allows us to implement feedback laws that read these electrochemical gradients and modify growth to enhance interface morphological stability.
}
    \label{fig:problem-setup}
\end{figure*}

To describe the characteristics of battery cycling and the formation of instabilities, we study the dynamics of anode-electrolyte interface as shown in
Fig.~\ref{fig:problem-setup} B.
The as-assembled flat interface begins to deform as instabilities grow during Lithium plating/deposition, while maintaining full contact for a period of time.
At high enough current densities (or long enough time), the incompatibility between the flat  electrolyte and the potentially curved surface of Lithium metal (due to inhomogeneous deposition) can cause delamination and void formation at anode-electrolyte interface (Fig.~\ref{fig:problem-setup} B).
As voids form, current localization accelerates plating in the contacted region and causes filaments to grow and become unstable \cite{eckhardt3DImpedanceModeling2022}.
To prevent void and filament growth, we consider the role of adding a thin interfacial material at the electrode-electrolyte interface that implements closed-loop feedback by coupling non-flat morphology to Lithium deposition/stripping. The added morphogenic interface (MINT) material, shown in green in Fig.~\ref{fig:problem-setup} B, could, under the right conditions, prevents contact loss and filament growth by implementing feedback between morphology and growth, which reduces the amplitude and increases the wavelength --- maintaining a  smoother and nearly flat interface --- of interface undulations. To understand the conditions under which this might be feasible, we first turn to a theoretical framework and determine a stability diagram for the dynamics of the MINT, and then consider different practial options for its implementation.

\section*{Model for morphogenic interface dynamics}





\begin{figure*} [t]
    \centering
    \includegraphics[width=\linewidth]{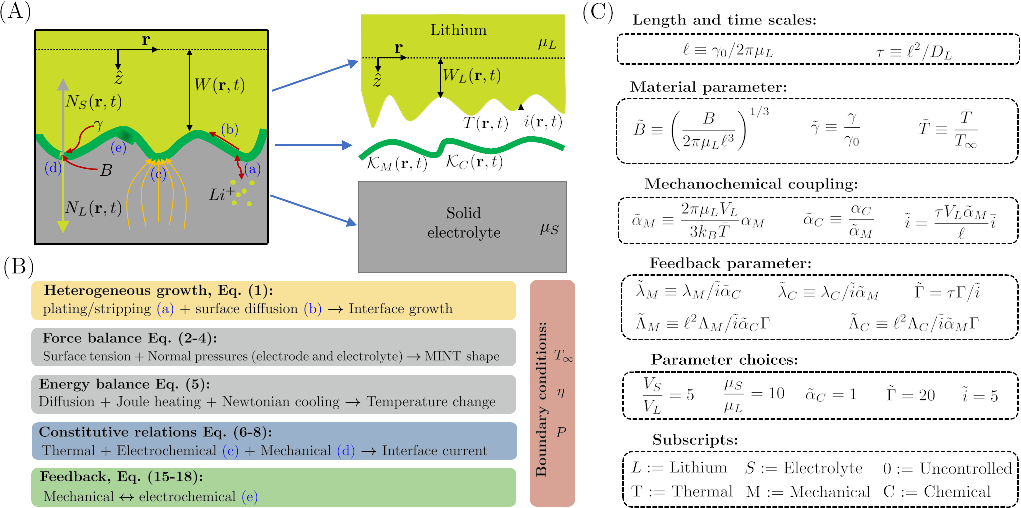}
    \caption{\textbf{Transport and feedback in a solid state battery}. (A) The anode interface will consist of lithium metal with a naturally curved boundary due to inhomogenieties in plating/stripping current density along the interface $i(\mathbf{r}, t)$, parameterized by the coordinates $\mathbf{r} = (x_1, x_2)$ with preferred normal displacement (along the $\hat{z}$ direction) given by $W_L(\mathbf{r}, t)$. Due to contact with a solid electrolyte that elastically prefers a flat boundary and a MINT layer with natural curvature $\mathcal{K}_M(\mathbf{r}, t)$, the interface normal displacement, $W(\mathbf{r}, t)$, adopts the value dictated by force balance \eqref{eq:normal-f-balance}. $N_S(\mathbf{r}, t), N_L(\mathbf{r}, t)$ are the normal pressures associate with deforming the solid electrolyte (with rigidity $\mu_S$) and lithium ($\mu_L$), while $B$ is the bending rigidity of the MINT and $\gamma$ is the effective surface tension of the electrode-MINT-electrolyte interface. Nonuniform current density $i(\mathbf{r}, t)$ also leads to variations in temperature along the interface, $T(\mathbf{r}, t)$. Here $\mathcal{K}_C(\mathbf{r}, t)$, which has units of inverse length, is related to lithium conductivity as given in \eqref{eq:echem-eta}. (B) A description of the equations defining the model that appear in the main text. The color matches the analogous part in Fig.~\ref{fig:problem-setup}C. Panel (C) shows compound variables and parameter values that are used throughout the text (see also SI table S2 for brief descriptions of these variables). 
    }
    \label{fig:model-diagram}
\end{figure*}

We assume that initially the electrode, electrolyte, and MINT are in full contact and the interface is nearly flat and coinciding with a horizontal plane characterized by coordinates $\mathbf{r} \equiv (x, y)$ and with a normal in the direction $\hat{z}$. The dynamic shape of the interface which varies during lithium stripping and plating is assumed to be given by $W_L(\mathbf{r},t)$ (Fig.~\ref{fig:model-diagram}), and is analogous to the rest configuration of a spring when completely unloaded. We assume a thin MINT layer, whose thickness is smaller than the wavelength of interface displacements controlling void formation ($\sim 10 \mu$m), and describe its mechanical behavior as a thin elastic shell with rest curvature $\mathcal{K}_M(\mathbf{r}, t)$ (which we will see later is a geometric field subject to feedback control in an engineered setting and can serve to stabilize the interface mechanically), bending rigidity $B$, and an effective surface tension for the combined electrode-MINT-electrolyte interface $\gamma$. 

The combined effects of electrokinetics associated with plating and stripping and the presence of the solid (elastic) electrolytes on either side of the interface causes it deform and move. Since elastic forces equilibrate much faster compared with deposition and diffusion timescales, the shape of the combined interface ${W}(\mathbf{r},t)$ is  determine by solving the equilibrium force balance equations --- that incorporates the effect of surface tension, MINT bending forces, and electrode-electrolyte normal indentation forces, denoted as $N_L(\mathbf{r}, t)$ and $N_S(\mathbf{r}, t)$ --- at each time step.  Mechanical forces and geometry of the combined interface will then modify the current density through mechanical and electrochemical feedback. In addition, current variation $i(\mathbf{r}, t)$ along the interface will lead to temperature variations $T(\mathbf{r}, t)$, which in turn modify the kinetics of deposition. For a complete list of variable and parameter definitions see Fig.~\ref{fig:model-diagram} and SI table S1.

\subsection*{Heterogeneous growth kinetics of the Lithium interface}

Plating or stripping of the lithium surface, is dictated by the imbalance in the current density and spatial diffusion, leading to an equation of motion for the interface given by
\begin{eqnarray}
   \underbrace{\partial_t {W}_{L}(\mathbf{r}, t)}_{\text{Interface growth}} = \underbrace{i(\mathbf{r}, t) V_{L}}_{\text{Plating/stripping}} + \underbrace{D_L \triangle W(\mathbf{r}, t)}_{\text{Self diffusion}},
   \label{eq:deposition} 
\end{eqnarray}
 where $V_{L}$ is the atomic volume of lithium, and the first term updates the shape of the lithium interface due to plating ($i(\mathbf{r}, t) > 0$) or stripping ($i(\mathbf{r}, t) < 0$) and the second term represents self (surface) diffusion of lithium with diffusivity $D_L$. 
 
\subsection*{Mechanical force balance}
 
 When ${W}_{L}(\mathbf{r}, t) \neq 0$ due to variations in the current density, its contact with the flat solid electrolyte (Fig.~\ref{fig:model-diagram}) leads to mechanical forces  that satisfy the following normal force balance equation\cite{landau1986theory}:
\begin{eqnarray}
     - \underbrace{B \triangle \left[\triangle W(\mathbf{r}, t) - \mathcal{K}_M(\mathbf{r}, t)\right]}_{\text{MINT bending}} + \underbrace{\gamma \triangle W(\mathbf{r}, t)}_{\text{Surface tension}}\nonumber\\  + \underbrace{N_L(\mathbf{r}, t)}_{\text{Electrode normal pressure}}  + \underbrace{ N_S(\mathbf{r}, t)}_{\text{Electrolyte normal pressure}}= 0, \label{eq:normal-f-balance}
\end{eqnarray}
where the first term is due to MINT bending relative to its rest natural curvature $\mathcal{K}_M(\mathbf{r}, t)$, the second is due to surface tension, and the last two terms are the normal pressures imposed by the electrode (subscript L) and electrolyte (subscript S) on the interface material that, for elastic substrates are given by the non-local terms associated with the Boussinesq-Cerruti solutions for the response of a half-space to surface forces \cite{love2013treatise}
\begin{eqnarray}
    N_L(\mathbf{r}, t) = \mu_L \int \frac{\triangle W(\mathbf{r}^\prime, t) - \triangle W_L(\mathbf{r}^\prime, t)}{|\mathbf{r} - \mathbf{r}^\prime|}  d^2r^{\prime}, 
    \label{eq:normal-electrode-force}
\end{eqnarray}
\begin{eqnarray}
     N_S(\mathbf{r}, t) = \mu_S \int \frac{\triangle W(\mathbf{r}^\prime, t) 
 }{|\mathbf{r} - \mathbf{r}^\prime|}  d^2r^{\prime}, 
     \label{eq:normal-electrolyte-force}
\end{eqnarray}
where $\mu_L, \mu_S$ are the elastic moduli of the electrode and electrolyte materials and $W_L(\mathbf{r}, t) \neq 0$ due to inhomogeneous deposition described by \eqref{eq:deposition}. 


\subsection*{Energy balance}
Since the interface dynamics are associated with heating which change the current, we need to consider how the temperature of the system changes as a function of joule heating and diffusion, leading to an equation for the system temperature $T({\bf r},t)$ given by
\begin{equation}
    \underbrace{\partial_t T(\mathbf{r}, t) }_{\text{Temp. change}}= \underbrace{D_T \triangle T(\mathbf{r}, t)}_{\text{Diffusion}} + \underbrace{\frac{i(\mathbf{r}, t) \eta}{c_T}}_{\text{Joule heating}} - \underbrace{\frac{T(\mathbf{r}, t) - T_{\infty}}{\tau_T}}_{\text{Newtonian cooling}}, \label{eq:temp}
\end{equation}
where $D_T$ is the thermal diffusion coefficient parallel to the interface,  $   i(\mathbf{r}, t) $ is the current through the interface with $\eta$ being a prescribed  surface over-potential (with units of energy), and $c_T$ is the  effective heat capacity (per unit area of the interface), and $\tau_T$ is the time-scale for thermal dissipation away from the interface to the bulk (which is assumed to be at temperature $T_{\infty}$). 



\subsection*{Constitutive relations}

Equations (\ref{eq:deposition}-\ref{eq:temp}) describe the effect of current on interface morphological, mechanical, and thermal properties, which will in turn modify the current. To characterize the current density, we use a modified Butler-Volmer equation that accounts for the effect of mechanical forces \cite{ahmad2017stability} and electrochemical transport \cite{carmona2023modeling}, so that we can write
\begin{equation}
    i(\mathbf{r}, t) = \bar{i}(T) \exp\left[  \underbrace{\alpha_{C} \frac{\eta_C(\mathbf{r}, t)}{k_B T}}_{\text{Electrochemical}} + \underbrace{\alpha_{M} \frac{\eta_M(\mathbf{r}, t)}{k_B T}}_{\text{Mechanical}}  \right], \label{eq:current}
\end{equation}
 where $\alpha_{C}$ and $\alpha_{M}$ are dimensionless constants, $\eta_M({\bf r},t)$ is the mechanical and $\eta_C({\bf r},t)$  is the chemical overpotential, and $k_BT$ is the usual Boltzmann factor. Furthermore, $\bar{i}(T) \equiv 2 i_{0} \sinh(\eta/{2 k_B T})$ is the unperturbed (flat interface) current density, which follows from the Butler-Volmer equation \cite{bard2022electrochemical}, with $i_0$ being the exchange current density and assuming a charge transfer coefficient of $0.5$. 
 
 The mechanical overpotential \cite{monroe2004interfacial, monroe2005elastic} is given by
\begin{eqnarray}
    \eta_{M}(\mathbf{r}, t) = V_{S} \left[-\frac{N_S(\mathbf{r}, t)}{3}\right] - V_{L} \left[\frac{N_L(\mathbf{r}, t)}{3}\right]
, \label{eq:pvchange}
\end{eqnarray}
where $[N_L(\mathbf{r}, t)/3]$, and $[-N_S(\mathbf{r}, t)/3]$ are the changes in pressure, relative to the flat state, in the electrode and electrolyte near the interface, and $V_{{S}}$ is the molar volume of lithium ions in the electrolyte. 
The electrochemical overpotential describes how chemical gradients and electric fields --- and therefore current through Fick's and Ohm's laws --- are modified due to interface curvature, which is described by the following equation [see Ref. \cite{carmona2023modeling} and SI for a derivation of these effects to linear order in interface curvature $\triangle W(\mathbf{r}, t)$]:  
\begin{eqnarray}
    \eta_{C}(\mathbf{r}, t) =  k_B T \left[h \mathcal{K}_C(\mathbf{r}^\prime, t) - \int \frac{\triangle W(\mathbf{r}^\prime, t) }{|\mathbf{r} - \mathbf{r}^\prime|}  d^2r^{\prime} \right], \label{eq:echem-eta}
\end{eqnarray}
 where $\mathcal{K}_C(\mathbf{r}, t)$, which has units of inverse length, and describes the position dependent interfacial conductivity of the electrode-MINT-electrolyte interface (which has thickness $h$), and is subject to feedback control using the MINT layer to potentially stabilize the interface. We note that the functional form of the electrochemical potential given by \eqref{eq:echem-eta} and the mechanical overpotential given by Eqs.~(\ref{eq:normal-electrode-force} - \ref{eq:normal-electrolyte-force}, \ref{eq:pvchange}) both have a nonlocal dependence on interface curvature. This completes the formulation of the governing equations for the thermo-electro-mechanical fields $W_L({\bf r},t),W({\bf r},t), T({\bf r},t)$ given by Eq. [1,2,5] subject to the closure relations [3-4,6-8].

\subsection*{Boundary and initial conditions}
 In the absence of feedback, $\mathcal{K}_M(\mathbf{r}, t) = 0$ and $\mathcal{K}_C(\mathbf{r}, t) = 0$, the system of equations given by Eqs.~(\ref{eq:deposition}-\ref{eq:echem-eta}) is closed and can be solved given the initial displacements $W\mathbf{r}, 0),W_L(\mathbf{r}, 0)$ and temperature $T(\mathbf{r}, 0)$. While the choice of boundary conditions does not affect the results we present below, for concreteness we assume
 \begin{eqnarray}
     \lim_{\mathbf{r} \to \infty} i(\mathbf{r}, t) = 0, \;\;\;\; \lim_{\mathbf{r} \to \infty} T(\mathbf{r}, t) = T_{\infty} \nonumber \\
     \lim_{\mathbf{r} \to \infty} W_L(\mathbf{r}, t) = 0, \;\;\;\; \lim_{\mathbf{r} \to \infty} W(\mathbf{r}, t) = 0. \label{eq:b-conds}
 \end{eqnarray}

\subsection*{Dimensional analysis}
Before we determine how mechanical properties and feedback mechanisms may control interface stability, we first consider the uncontrolled case, i.e. with vanishing rest curvature and conductance of the MINT, defined by the absence of the MINT layer and temperature variations: 
\begin{eqnarray}
     B = 0, D_L = D_{0}, \; \mathcal{K}_M = \mathcal{K}_C = 0, \;\gamma = \gamma_0, \; \bar{T} = T_{\infty}, \label{eq:uncontrolled-params}
\end{eqnarray} 
where $\gamma_0$ is the surface tension between the electrode and electrolyte and $D_{0}$ is the surface diffusivity of lithium in the absence of the MINT layer. This allows us to 
use dimensional analysis to characterize the parameters in the problem, as depicted in Fig.~\ref{fig:problem-setup} C. Using the surface tension between the electrode and electrolyte $\gamma_0$ and the lithium rigidity $\mu_L$, we obtain the length scale $\ell \equiv \gamma_0/2 \pi \mu_L$. Using this length scale and the lithium (surface) diffusivity $D_L$, we have the time scale $\tau \equiv \ell^2/D_L$. While the parameters $\alpha_C$ and $\alpha_M$, which give the strength of the electrochemical and mechanochemical couplings respectively, are dimensionless, we defined the normalized constants $\tilde{\alpha}_C$ and $\tilde{\alpha}_M$ (see Fig.~\ref{fig:problem-setup} C) that contain the parameter combinations appearing in subsequent calculations. To normalize the average current density $\bar{i}$, which appears in \eqref{eq:current}, we use the length and time scales defined above to obtain a velocity and the reciprocal of the molar volume of lithium $V_L$ to define a density, giving the normalized current $\tilde{i}$ (see Fig.~\ref{fig:problem-setup} C). In addition, we also normalized the material parameters using their reference values in \eqref{eq:uncontrolled-params} to get the normalized bending rigidity of the MINT $\tilde{B}$, the normalized surface tension $\tilde{\gamma}$, and normalized temperature $\tilde{T}$ (see Fig.~\ref{fig:problem-setup} C). 

\section*{Stability of morphogenic interface}

{Rather than solve equations~(\ref{eq:deposition}-\ref{eq:b-conds}) for specific initial and boundary conditions, here we consider the stability of an initially uniform interface, which is described by a constant current density $\bar{i}$ (bar represents average quantities), temperature $\bar{T}$, and displacement $\bar{W} = {W}_L = \bar{i} V_L t$. Expanding these four fields to linear order around this homogeneous base state, we carry out a linear stability  analysis in terms of the spatial Fourier transform, which for a quantity $F(\mathbf{r}, t)$} is defined as
$\hat{F}(\mathbf{q}, t) \equiv \int d^2 \mathbf{r} \;F(\mathbf{r}, t) e^{-i \mathbf{q} \cdot \mathbf{r}}$.
For each wavenumber $\mathbf{q}$, we calculate the growth rate $\sigma(\mathbf{q})$ of the interface, which grows according to 
\begin{eqnarray}
    \hat{W}(\mathbf{q},t) = \hat{W}(\mathbf{q}, 0) \;\;e^{\sigma(\mathbf{q}) t}. \label{eq:growth-rate}
\end{eqnarray}
This allows us to determine the wavenumber $\mathbf{q}^*$ with the maximum value of the growth rate $\sigma^* \equiv \sigma(\mathbf{q}^*)$; a change in the sign of $\sigma^*$ will characterize the transition from a stable to an unstable interface as a function of the parameters in the problem. We denote the maximum growth rate and corresponding wavenumber of the uncontrolled interface as 
\begin{eqnarray}
    \sigma^*_0 \equiv \max_{\tilde{q}}  \sigma_0(\tilde{q}), \;\;\;\;\;    \tilde{q}^*_0 \equiv \underset{\tilde{q}}{\mathrm{argmax}}\; \sigma_0(\tilde{q}), \label{eq:uncontrolled-max}
\end{eqnarray}
which will serve as reference values for comparison as the parameters of the systems are changed from those given in \eqref{eq:uncontrolled-params}.


\subsection*{Uncontrolled Charging / Discharging} \label{uncontrolled}

To obtain the growth rates $\sigma(\mathbf{q})$ in the following analysis (see SI for more details including for the controlled case), we first use \eqref{eq:normal-f-balance}, Fourier transformed as given in the SI, to solve for $\hat{W}_L(\mathbf{q},t)$ in terms of $\hat{W}(\mathbf{q},t)$, using the expressions of the normal forces Eqs.~(\ref{eq:normal-electrode-force}-\ref{eq:normal-electrolyte-force}). Then, using the expressions for the mechanical and electrochemical overpotentials, given in Eqs.~(\ref{eq:pvchange}-\ref{eq:echem-eta}), the current can be eliminated in terms of $\hat{W}(\mathbf{q},t)$ using \eqref{eq:current}. The results for $\hat{W}_L(\mathbf{q},t)$ and $\hat{i}(\mathbf{q},t)$ in \eqref{eq:deposition} are then used to obtain a first order differential equation in time, from which we can read off the (wavelength-dependent) growth rate $\sigma(\mathbf{q})$ defined in \eqref{eq:growth-rate}.

Using the normalized average current $\Tilde{i}$ (see Fig.~\ref{fig:model-diagram} C), the ratio of molar volumes $\Tilde{V}$, ratio of elastic moduli $\Tilde{\mu}$, and the length scale $\ell$, the normalized wavenumber $\tilde{q} \equiv \ell |\mathbf{q}|$, and time scale $\tau \equiv \ell^2/D_{0}$, we write the growth rate in the uncontrolled case as (see SI-S2  for details of the calculation) 
\begin{eqnarray}
    \sigma_0(\tilde{q}) = \frac{-\tilde{q} + \tilde{i}  \left[\tilde{\alpha}_C- \tilde{q} + \tilde{\mu} (\tilde{V} - 1)\right]}{1 + \tilde{q} + \Tilde{\mu}} \frac{\tilde{q}}{\tau}. \label{eq:uncontrolled-rate}
\end{eqnarray}
Since typically $\Tilde{V} > 1$ \cite{mistry2020molar}, the growth rate given in \eqref{eq:uncontrolled-rate} is always positive for small wavenumbers (long wavelengths) and will grow with the applied current according to: 
\begin{eqnarray}
     \tau \sigma_0(\tilde{q}) = \frac{\tilde{i} \tilde{q}}{\tau (1 + \Tilde{\mu})}  \left[\tilde{\alpha}_C + \tilde{\mu} (\tilde{V} - 1)\right], \;\;\text{as}\;\; \Tilde{q} \to 0 \label{eq:uncontrolled-zero-q}
\end{eqnarray}
Thus, this system is unstable for any value of current, due to two sources of instability: One mechanical ($\Tilde{V} > 1$) and the other electrochemical ($\alpha_C > 0$). The mechanical instability results since negatively curved regions of the interface ---corresponding to a lithium bumps growing into the electrolyte --- lead to increased pressure near the interface which in turn drive more lithium ions towards the lithium metal when $\Tilde{V} > 1$, amplifying the growth of the lithium into the electrolyte \cite{mistry2020molar}. The electrochemical instability happens due to the coupling of the curvature to current through \eqref{eq:echem-eta}, which can be understood intuitively since a lithium bump (negative interface curvature) would lead to an amplification of the local electric fields, which in turn leads to an increase in current density that exacerbates the instability \cite{carmona2023modeling}.

\begin{figure*}[t]
    \centering
    \includegraphics[width=\linewidth]{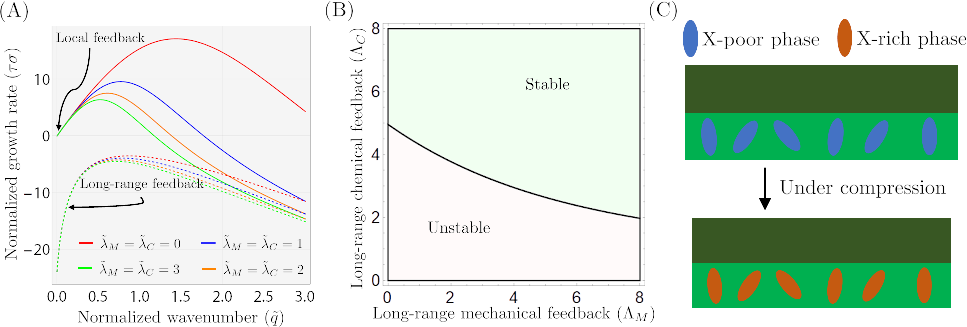}
    \caption{\textbf{Stability diagram for uncontrolled and controlled morphogenic interfaces}. (A) Growth rate as a function of wavenumber for different choices of the local and long-range feedback model, Eqs.~(\ref{eq:nonlocal-1}- \ref{eq:nonlocal-2}). The parameter definitions and values used in this figure are given in Fig.~\ref{fig:model-diagram} C and SI table.~S1. See also \eqref{eq:uncontrolled-rate} and SI text for details of how we calculate the growth rate $\sigma(\mathbf{q})$. Solid curves give the growth rate of the interface, for local feedback only ($\Tilde{\Lambda}_M = \Tilde{\Lambda}_C = 0$), while dashed curves give the growth rate for long-range feedback ($\Tilde{\Lambda}_M = \Tilde{\Lambda}_C = 3$). (B) Stability phase diagram, where stable corresponds to regions in parameter space where $\sigma(\mathbf{q}) < 0, \forall q$, while the unstable region violates this condition. This shows that increasing the long-range feedback parameters, $\Tilde{\Lambda}_M$ and $\Tilde{\Lambda}_C$ increases flat interface stability. The local feedback parameters are fixed at $\Tilde{\lambda}_M = \Tilde{\lambda}_C = 2$ and other model parameter values are given in Fig.~\ref{fig:model-diagram} C. (C) An example showing how long-range mechanical and chemical feedback can be realized. The symbol X here represent an atom or a molecule (e.g., sodium or potassium), which we term the morphogen in analogy to biological morphogenesis. The transition from the X-poor to the X-rich phase is triggered by compression, which in turn depends on the interface morphology (Fig.~\ref{fig:model-diagram}). During a transition, the inclusion function as sources or sinks of X, which will diffuse towards or away from surrounding regions, affecting Lithium diffusivity in the case of electrochemical feedback and swelling in the case of mechanical feedback. 
    }
    \label{fig:phase-diagram}
\end{figure*}

\section*{Feedback control to stabilize the MINT}

We see that in the absence of any regulatory process, there is a tendency for the interface to become unstable to long wavelength fluctuations, where elastic effects To stabilize the dynamics of the electrochemical interface, we now consider the role of the morphogenic interface which can be engineered to explicitly modify the mechanical and chemical driving forces on the interface, via potential closed-loop feedback mechanisms. We model such effects through the dynamic response of the mechanical and chemical natural curvatures, $\mathcal{K}_M(\mathbf{r}, t)$ and $\mathcal{K}_C(\mathbf{r}, t)$. 
Translation and rotation invariance (associated with the assumption that the MINT is so thin as to be not affected by the battery size and shape) imply that $\mathcal{K}_M(\mathbf{r}, t)$ and $\mathcal{K}_C(\mathbf{r}, t)$ can depend on $W(\mathbf{r}, t)$ only through the interface curvature, which is given by $\triangle W(\mathbf{r}, t)$ to linear order. Thus, the most general forms of linear feedback can be written as:
\begin{eqnarray}
    \underbrace{\mathcal{K}_M(\mathbf{r}, t)}_{\text{MINT curvature}}  = \int_{-\infty}^t \underbrace{\mathcal{G}_M(\mathbf{\mathbf{r} - \mathbf{r} ^\prime}, t - t^\prime)}_{\text{feedback kernel}} \underbrace{ \triangle W(\mathbf{r} ^\prime, t^\prime)}_{\text{curvature}} dt^\prime d^2\mathbf{r}^\prime,  \label{eq:swelling}
\end{eqnarray}
\begin{eqnarray}
    \underbrace{\mathcal{K}_C(\mathbf{r}, t)}_{\text{MINT conductance}}  =  \int_{-\infty}^t \underbrace{\mathcal{G}_C(\mathbf{\mathbf{r} - \mathbf{r}^\prime}, t - t^\prime)}_{\text{feedback kernel}} \underbrace{ \triangle W(\mathbf{r} ^\prime, t^\prime)}_{\text{curvature}} dt^\prime d^2\mathbf{r}^\prime.  \label{eq:conductance}
\end{eqnarray}
where the kernels $\mathcal{G}_M(\mathbf{\mathbf{r} - \mathbf{r} ^\prime}, t - t^\prime)$ and $\mathcal{G}_C(\mathbf{\mathbf{r} - \mathbf{r} ^\prime}, t - t^\prime)$  describes the effect of interface curvature at position $\mathbf{r}^\prime$ and time $t^\prime$ on the MINT mechanical rest curvature and conductance at a later time $t$ and position $\mathbf{r}$.  To allow for the regulation of undulations with both small and long wavelengths,  the kernels can in general include both local and long-range feedback in space-time --- such feedback may be mediated by diffusing signals as in the case of growing leaves~\cite{al2022grow}.


To study the effect of closed-loop feedback, we consider a diffusing signal that travels fast compared to the Lithium diffusive timescale $\tau$ so that a steady state approximation is valid. Then, we may write
\begin{eqnarray}
   \mathcal{G}_{M}(\mathbf{r}, t) = - \left[ \underbrace{\lambda_M \delta(\mathbf{r}) \delta(t)}_{\text{Local}} -  \underbrace{\frac{\Lambda_M}{2 \pi} \log\left(\frac{|\mathbf{r}|}{r_0}\right)  e^{- \Gamma t}}_{\text{Long-range}} \right],  \label{eq:nonlocal-1}\\
   \mathcal{G}_{C}(\mathbf{r}, t) = - \left[ \underbrace{\lambda_C \delta(\mathbf{r}) \delta(t)}_{\text{Local}} -  \underbrace{\frac{\Lambda_C}{2 \pi} \log\left(\frac{|\mathbf{r}|}{r_0}\right)  e^{- \Gamma t}}_{\text{Long-range}} \right], \label{eq:nonlocal-2}\;\;\;
\end{eqnarray}
where $\lambda_M,\lambda_c$ are constants defining the strength of spatio-temporally local feedback, while $\Lambda_M$, $\Lambda_C$ are constants representing the strength of the non-local feedback. The function $\log(|\mathbf{r}|/r_0)$ is the steady state solution of the diffusion equation with a point source at the origin and $r_0$ being a short distance cutoff, which approximates the effect of a diffusing agent --- morphogen in the biological context \cite{turingChemicalBasisMorphogenesis1952, wolpert1969positional}. The exponential function in time, with $\Gamma >0$, is associated with the finite duration over which the signal (diffusing agent, or morphogen) is produced in response to nonzero interface curvature. With the feedback laws given in Eq.~(\ref{eq:nonlocal-1}-\ref{eq:nonlocal-2}), the system now can be stabilized, even for very long wavelengths, with the growth rate
\begin{eqnarray}
     \tau \sigma_0(\tilde{q}) = -\frac{\tilde{i} \tilde{\Gamma}}{2} \left( 1  - \sqrt{1 - \frac{4 \tilde{B} \tilde{i} \tilde{\Lambda}_C}{(1 +  \tilde{\mu}) \tilde{\Gamma}}}\right), \;\;\text{as}\;\; \Tilde{q} \to 0, \label{eq:uncontrolled-zero-q}
\end{eqnarray}
which is negative when the feedback parameter $\tilde{\Lambda}_C$, defined in Fig.~\ref{fig:problem-setup} C, is positive. 

We note that the exact choice of $\mathcal{G}_{M}(\mathbf{r}, t)$ and $\mathcal{G}_{C}(\mathbf{r}, t)$ is not essential to achieve complete stability, as long as it is sufficiently long-range. In Fourier space, the long-range requirement can be stated precisely: Stability at $\tilde{q} \to 0$ requires that in this limit $\hat{\mathcal{G}}_{C}(\mathbf{q}, t) \sim 1/\tilde{q}^{p}$, with $p \geq 1$, and similarly for $\hat{\mathcal{G}}_{M}(\mathbf{r}, t)$, with $p \geq 3$.

\subsection*{Decreasing growth rate with MINT material properties}

Having established the uncontrolled growth as a reference case in \eqref{eq:uncontrolled-zero-q}, we consider how adding a morphogenic interface will modify the mechanical and chemical properties of the interface and stabilize the flat state --- which we quantify as a reduction in the growth rate $\sigma^*$ relative to $\sigma^*_0$. We considered the effect of three internal parameters (surface diffusion $\Tilde{D}_L$, surface tension $\Tilde{\gamma}$, and MINT bending rigidity $\Tilde{B}$), in addition to the role of external temperature and pressure. 

Increasing the surface diffusion coefficient $\tilde{D}_L$ is thought to stabilize flat interfaces in solid state batteries \cite{jackle2018self} by redistributing lithium atoms from peaks to valleys in the interface. However due to the nonlocality of the sources of instability, Eqs.~(\ref{eq:normal-electrode-force}- \ref{eq:normal-electrolyte-force}) and \eqref{eq:echem-eta}, the instability always dominates for small wavenumbers, $\Tilde{q} \ll 1$. In fact, as shown in SI Fig.~S3, the slope of $\sigma(\Tilde{q})$ at $\Tilde{q} = 0$ is unchanged from its value in the uncontrolled case, which is given in \eqref{eq:uncontrolled-zero-q}. However, increasing $\tilde{D}_L$ decreases the maximum growth rate $\sigma^*$ (SI Fig.~S3 B). 

As shown in Fig.~\ref{fig:phase-diagram} A, (see also SI Fig.~S3 C-D), increasing MINT rigidity $\tilde{B}$ and surface tension $\tilde{\gamma}$ tends to decrease $\sigma^*$ by providing mechanical support, which reduces the pressure variations across the interface that lead to the mechanochemical instability described in the previous section. Just as in the case of surface diffusion, these mechanical stabilizing factors act locally, as given in \eqref{eq:normal-f-balance}, and therefore do not modify the $\Tilde{q} \to 0$ behavior of $\sigma(\Tilde{q})$. These results support the use of adding rigid interlayers such as Al$_2$O$_3$ and modulating impurities in the solid electrolyte to increase the net surface tension, while still maintaining wetting of lithium onto the electrolyte as described in \cite{sharafi2017surface}.

Since temperature affects interfacial resistance and other properties of the interface, the dynamics of heat flow may influence interface stability. Increasing temperature due to high applied current can lead to self-healing of lithium dendrites \cite{li2018self}. In addition, we consider here the role of local temperature increases due to current hot spots in suppressing interface undulations. 
Our analysis (see SI text and SI Fig.~S3) shows that the stabilizing effect of temperature is more pronounced for slow thermal diffusion $\tau_T D_T \ll \ell^2$, long cooling timescale $\tau_T$ over which heat escapes from the interface, and small volumetric heat capacity $c_T$ that amplifies temperature increase for the same amount of applied current density. 

Lastly, to leading order in our analysis, pressure does not affect the growth rate $\sigma^*$ since \eqref{eq:pvchange} only depends on the changes in pressure relative to the flat state \cite{monroe2004interfacial, monroe2005elastic, ahmad2017stability, mistry2020molar}. However it is expected that pressure will increase stability of an interface due to higher order effects, such as plastic deformation \cite{barai2017lithium, shishvan2021initiation}, which are not accounted for in our linear analysis. 


\section*{Discussion}

Inspired by regulated morphogenesis in biological systems, we have explored how to create self-regulating electrochemical interfaces (MINTs) with  feedback control mechanisms that sense deviations from and restore a flat interface morphology. Since the sources of interface instability lead to long wavelength instabilities in the uncontrolled case, it is not surprising that only long-range feedback can truly lead to a stable interface with negative growth rate (Fig.~\ref{fig:phase-diagram} A-B). 



Closed-loop feedback requires MINT materials that are capable of sensing the state of the interface (morphology, stress, concentrations, etc.,) and modifying its properties (e.g., conductivity, swelling, etc.,) accordingly. One such mechanism for sensing interface morphology is through inclusions inside the MINT is (shown in Fig.~\ref{fig:phase-diagram} C) that undergo a diffusive phase transition into an X-rich phase (e.g. X=sodium for example)  when subjected to compression. By placing the inclusions on the side of the MINT closer to the electrode, they will transition to the X-rich phase only for when the interface curvature is negative (which results from a lithium bump pushing into the electrolyte), thus effectively sensing morphology.  Mechanical actuation can be realized by coupling the amount of swelling (which determines the MINT natural curvature $\mathcal{K}_M$) to the concentration of X (morphogen) signal outside the inclusions. Another possibility is using electrochemical feedback which can be realized by coupling the conductivity of the MINT to the amount of X, for example, sodium may reduce deposition by occupying sites important for lithium diffusion. Lastly, the long-range nature of the feedback can be accomplished through diffusion of X through the MINT, with the inclusions functioning either as sources or sinks depending on the direction of the phase transition (to or from the X-rich phase).

Overall, our study provides a benchmark for future studies of self-regulating MINTs, and our predictive model might help guide the discovery of promising materials that stabilize electrochemical chemical interfaces for energy intensive applications.

\matmethods{ }


\acknow{This work was supported by the Defense Advanced Research Project Agency (DARPA) under contract no. HR00112220032. The results contained herein are those of the authors and should not be interpreted as necessarily representing the official policies or endorsements, either expressed or implied, of DARPA, or the U.S. Government.}

\showacknow{} 

\bibliography{references}

\end{document}